\documentclass[twocolumn]{aastex63}
\usepackage{stfloats}
\usepackage{float}
\usepackage{graphicx}
\usepackage{natbib}
\usepackage{amsmath}
\usepackage{autobreak}
\usepackage{multirow}
\usepackage{makecell}
\usepackage{color}
\usepackage{bm}
\defcitealias{casa2018}{CV18}
\defcitealias{wei2018}{WEI18}
\defcitealias{maw2018}{MAW18}

\begin{document}

\newcommand{\vdag}{(v)^\dagger}
\newcommand\aastex{AAS\TeX}
\newcommand\latex{La\TeX}

\received{}
\revised{}
\accepted{}
\submitjournal{ApJ}

\shorttitle{Correction to the \textit{Gaia} DR2 photometric colors}
\shortauthors{Niu et al.}
\graphicspath{{./}{figures/}}


\title{Correction to the photometric colors of the Gaia Data Release 2 with the stellar color regression method}

\correspondingauthor{Haibo Yuan}
\email{yuanhb@bnu.edu.cn}

\author{Zexi Niu}
\affiliation{National Astronomical Observatories,
Chinese Academy of Sciences\\
20A Datun Road, Chaoyang District,
Beijing, China}

\author{Haibo Yuan}
\affiliation{Department of astronomy,
Beijing Normal University \\
19th Xinjiekouwai Street, Haidian District,
Beijing, China}

\author{Jifeng Liu}
\affiliation{National Astronomical Observatories,
Chinese Academy of Sciences\\
20A Datun Road, Chaoyang District,
Beijing, China}

\begin{abstract}
The second \textit{Gaia} data release (DR2) delivers accurate and homogeneous photometry data of the whole sky to an exquisite quality, reaching down to the unprecedented milli-magnitude (mmag) level for the $G$, $G_{\rm RP}$, and $G_{\rm BP}$ passbands. However, the presence of magnitude-dependent systematic effects at the 10 mmag level limits its power in scientific exploitation. In this work, using about half-million stars in common with the LAMOST DR5, we apply the spectroscopy-based stellar color regression method to calibrate the \textit{Gaia} $G-G_{\rm RP}$ and $G_{\rm BP}-G_{\rm RP}$ colors. With an unprecedented precision of about 1 mmag, systematic trends with $G$ magnitude are revealed for both colors in great detail, reflecting changes in instrument configurations. Color dependent trends are found for the $G_{\rm BP}-G_{\rm RP}$ color and for stars brighter than $G$ $\sim$ 11.5 mag. The calibration is up to 20 mmag in general and varies a few mmag/mag. A revised color-color diagram of \textit{Gaia} DR2 is given, and some applications are briefly discussed.

\end{abstract}



\section{Introduction} \label{sec:intro}

The second data release of the \textit{Gaia} mission provides photometry data in the $G$ band for approximately 1.7 billion sources and in the integrated $G_{\rm BP}$ and $G_{\rm RP}$ bands for approximately 1.4 billion sources calibrated to a consistent and homogeneous photometric system\citep{gaia2016,gaia2018}. The $G$ magnitude is measured in the astrometric field and extracted by a point-spread-function fitting with a typical formal uncertainty under 1 mmag for stars of  $G$ $<$ 16 mag. The $G_{\rm BP}$ and $G_{\rm RP}$ magnitudes are obtained by the $BP$ and $RP$ photometer with larger formal uncertainties, about a few mmag for stars of $G$ $<$ 15 mag\citep{evans2018,riello2018}. With such an enormous data volume and exquisite data quality, \textit{Gaia} is able to present a significant advance in all photometric investigations.

To make full use of the mmag precision photometry yielded by the \textit{Gaia} DR2, photometric calibration to mmag precision is required. However, the systematic effects at the 10 mmag level or the higher are shown by a number of photometric validation tests. \citet{arenou2018} firstly show a global increase ($ \sim $2 mmag/mag) of the $G-G_{\rm BP}$ residuals with increasing $G$ magnitude, by subtracting fitted $G-G_{\rm BP}$ as a 3-order polynomial function of $G_{\rm BP}-G_{\rm RP}$. They also make comparisons with the external catalogues and find a similar global variation up to 10 mmag/mag. Other further studies, using well-calibrated, high-quality spectral libraries, \citeauthor*{casa2018} (\citeyear{casa2018}, hereafter CV18), \citeauthor{wei2018} (\citeyear{wei2018}, hereafter WEI18), and \citeauthor*{maw2018} (\citeyear{maw2018}, hereafter MAW18) conduct synthetic photometry and compare it with the \textit{Gaia} DR2 photometry. All of them detect a similar tendency in the $G$ magnitude and propose a linear correction for 3.4, 3.5, and 3.2 mmag/mag separately. Moreover, \citetalias{wei2018} and \citetalias{maw2018} find a significant color-dependent offset of 20 mmag between faint and bright stars in $G_{\rm BP}$. Nevertheless, since only hundreds of stars are available in their method, their linear corrections have the exact tendency yet lack of fine details.

Taking advantage of the fact that millions of stellar spectra and their associated precise stellar atmospheric parameters are now available, \citet{yuan2015} has proposed a spectroscopy-based stellar color regression (SCR) method to achieve mmag precision color calibration of wide-field imaging surveys. The method was applied to the Sloan Digital Sky Survey (SDSS; \citealt{sdss}) Stripe 82 data, achieving a precision $\sim$ 2 -- 5 mmag for different SDSS colors. It is straightforward and able to capture subtle structures thanks to the intensive database. Analogous to the SDSS database, the Large Sky Area Multi-Object Fiber Spectroscopic Telescope (LAMOST; \citealt{zhao2012,deng2012,liu2014}) spectroscopic sky survey is also appropriate to do so. It begins in 2012, acquiring thousands of stellar spectral per night with the spectral resolution R $\sim$ 1800 and accumulating more than 8 million stellar spectra in DR5. Effective temperature $T_{\rm eff}$, surface gravity log $g$, and metallicity [Fe/H], are delivered from the spectra through the \uppercase{LAMOST} Stellar Parameter pipeline (LASP; \citealp{luo2015,wu2011}) with the precision of about 110 K, 0.2 dex, and 0.1 dex, respectively \citep{luo2015}. In this work, combing the LAMOST DR5 spectroscopy and the \textit{Gaia} DR2 photometry, we apply the SCR method and reach new corrections to the \textit{Gaia} photometry at an unprecedented precision of about 1 mmag. 

The paper is organized as follows. We describe our data selection in Section \ref{sec:data}. The method and result are demonstrated in Section \ref{sec:methodresult}. The result validation and the comparison with previous researches are discussed in Section \ref{sec:dis}. We summarize our paper in Section \ref{sec:sum}.

\section{Data} \label{sec:data}

\begin{figure}[htbp] 
    \centering 
    \includegraphics[width=3.5in]{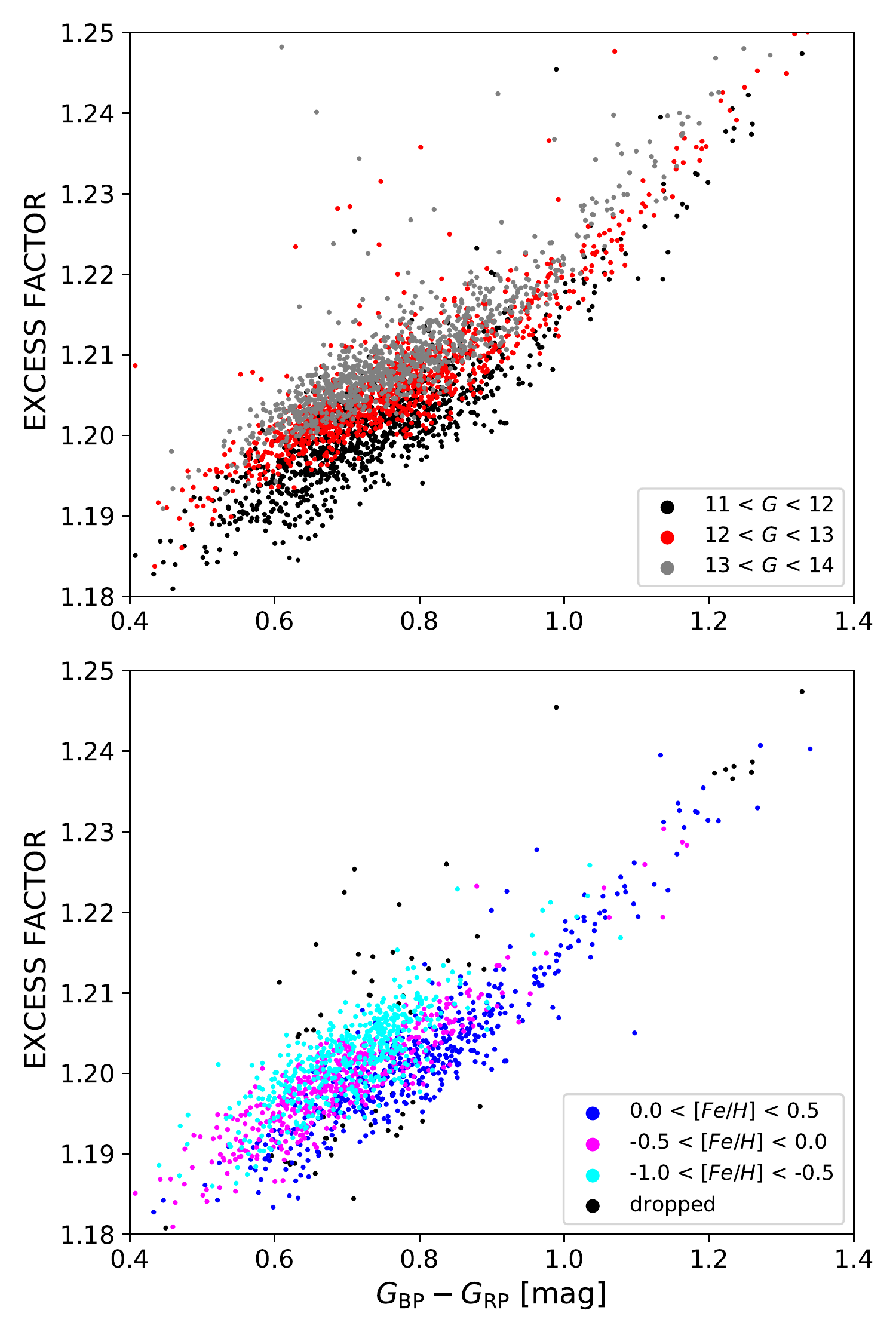} 
    \caption{\textit{Top}: \textit{Gaia} DR2 observed $G_{\rm BP}-G_{\rm RP}$ color against excess factor for three different $G$ magnitude bins at 1 mag interval. Five hundred randomly selected stars are plotted for each bin. \textit{Bottom}: same plot but for 3 metallicity bins of the same $G$ magnitude range 11-12 mag. 
    The black dots are dropped suspicious objects in the 2$\sigma$-clipping process. \label{fig:fig1.1}} 
\end{figure}

Stars having high-precision photometric colors and well-determined stellar atmospheric parameters are required in our study. Here we consider the following constraints to guarantee the data quality: 

\begin{itemize}
   \item [1)]
   \texttt{phot}$\_$\texttt{bp}$\_$\texttt{mean}$\_$\texttt{flux}$\_$\texttt{over}$\_$\texttt{error} $>$ 100
    \item [2)]
    \texttt{phot}$\_$\texttt{g}$\_$\texttt{mean}$\_$\texttt{flux}$\_$\texttt{over}$\_$\texttt{error} $>$ 100
  \item [3)]
\texttt{phot}$\_$\texttt{rp}$\_$\texttt{mean}$\_$\texttt{flux}$\_$\texttt{over}$\_$\texttt{error} $>$ 100
\item [4)]
\texttt{phot}$\_$\texttt{proc}$\_$\texttt{mode} = 0
\item [5)]
\texttt{duplicated}$\_$\texttt{source} = \texttt{False}
\item [6)]
\texttt{phot}$\_$\texttt{bp}$\_$\texttt{rp}$\_$\texttt{excess}$\_$\texttt{factor} (see text for limit)
\item [7)]
$E(B–V)_{\rm SFD} <$ 0.05 mag
\item [8)]
$|\texttt{glat}|$ $>$ 20 $\deg$
\item [9)]
$|Z|\ >$ 0.2 kpc, where Z is vertical distance to the Galactic disk
\item [10)]
the signal-to-noise ratios for the g band ($S/N_{\rm g}$) of the LAMOST spectra are larger than 30
\item [11)]
$T_{\rm eff} >$ 4500 K for dwarf stars
\item [12)]
remove variable stars
\end{itemize}  

Among them, criteria (1)-(6) are applied to the \textit{Gaia} DR2 data. \texttt{phot}$\_$\texttt{bp}$\_$\texttt{rp}$\_$\texttt{excess}$\_$\texttt{factor} is defined as the excess of flux in the $G_{\rm BP}$ and $G_{\rm RP}$ with the respect to the $G$ band, which is believed to be caused by background and contamination issues affecting the $G_{\rm BP}$ and $G_{\rm RP}$ data \citep{evans2018}. We find that it has a significant systematic trend with apparent $G$ magnitude and [Fe/H]. As shown in the Figure \ref{fig:fig1.1}, the \texttt{phot}$\_$\texttt{bp}$\_$\texttt{rp}$\_$\texttt{excess}$\_$\texttt{factor} systematically increases with the decreasing $G$ magnitude (about 0.005 per magnitude) and decreases with increasing metallicity (about 0.001 per 0.5 dex) within a given $G$ magnitude range. We thus bin our sample in the $G$ magnitude and [Fe/H] space with 1 mag and 0.5 dex intervals respectively and apply a 2$\sigma$-clipping on the \texttt{phot}$\_$\texttt{bp}$\_$\texttt{rp}$\_$\texttt{excess}$\_$\texttt{factor} to remove outliers in each bin (see Figure \ref{fig:fig1.1}). About 4$\%$ of objects are excluded in this step. Note that due to the relatively large widths of the magnitude and [Fe/H] bins, the estimated $\sigma$ values are larger than the scatters caused by the errors of the \texttt{phot}$\_$\texttt{bp}$\_$\texttt{rp}$\_$\texttt{excess}$\_$\texttt{factor}.
Therefore, a 2$\sigma$-clipping is used, a little bit excessive but secure.

We use the \citeauthor*{sfd} (\citeyear{sfd}, hereafter SFD) dust reddening map and criteria (7)-(9) to select low extinction stars for reliable reddening correction. 

The empirically determined, temperature and reddening dependent reddening coefficients R($G-G_{\rm RP}$) and R($G_{\rm BP}-G_{\rm RP}$) are given by the following functions (Sun et al., to be submitted):

\begin{equation}\label{eq1}
 \begin{aligned}
  &  R(G_{\rm BP}-G) = 0.901 - 0.225 \times E(B-V)_{\rm SFD} + 0.178 \\
  &  \times {E(B-V)}^2_{\rm SFD} - 6.422 \times 10^{-5} \times T_{\rm eff} - 3.982 \times 10^{-7} \\
  &  \times T_{\rm eff}  \times E(B-V)_{\rm SFD} + 3.541 \times 10^{-9} \times {T_{\rm eff}}^2
 \end{aligned} 
\end{equation}

\begin{equation}\label{eq2}
 \begin{aligned}
  &  R(G_{\rm BP}-G_{\rm RP}) = 1.060  - 0.253 \times E(B-V)_{\rm SFD} + 0.264 \\
  &  \times {E(B-V)}^2_{\rm SFD} + 5.535 \times 10^{-6} \times T_{\rm eff} -4.505\times 10^{-5} \\
  &  \times T_{\rm eff} \times E(B-V)_{\rm SFD} + 6.64 \times 10^{-9} \times {T_{\rm eff}}^2
 \end{aligned}
\end{equation}

\begin{figure}[htbp] 
    \centering 
    \includegraphics[width=3.5in]{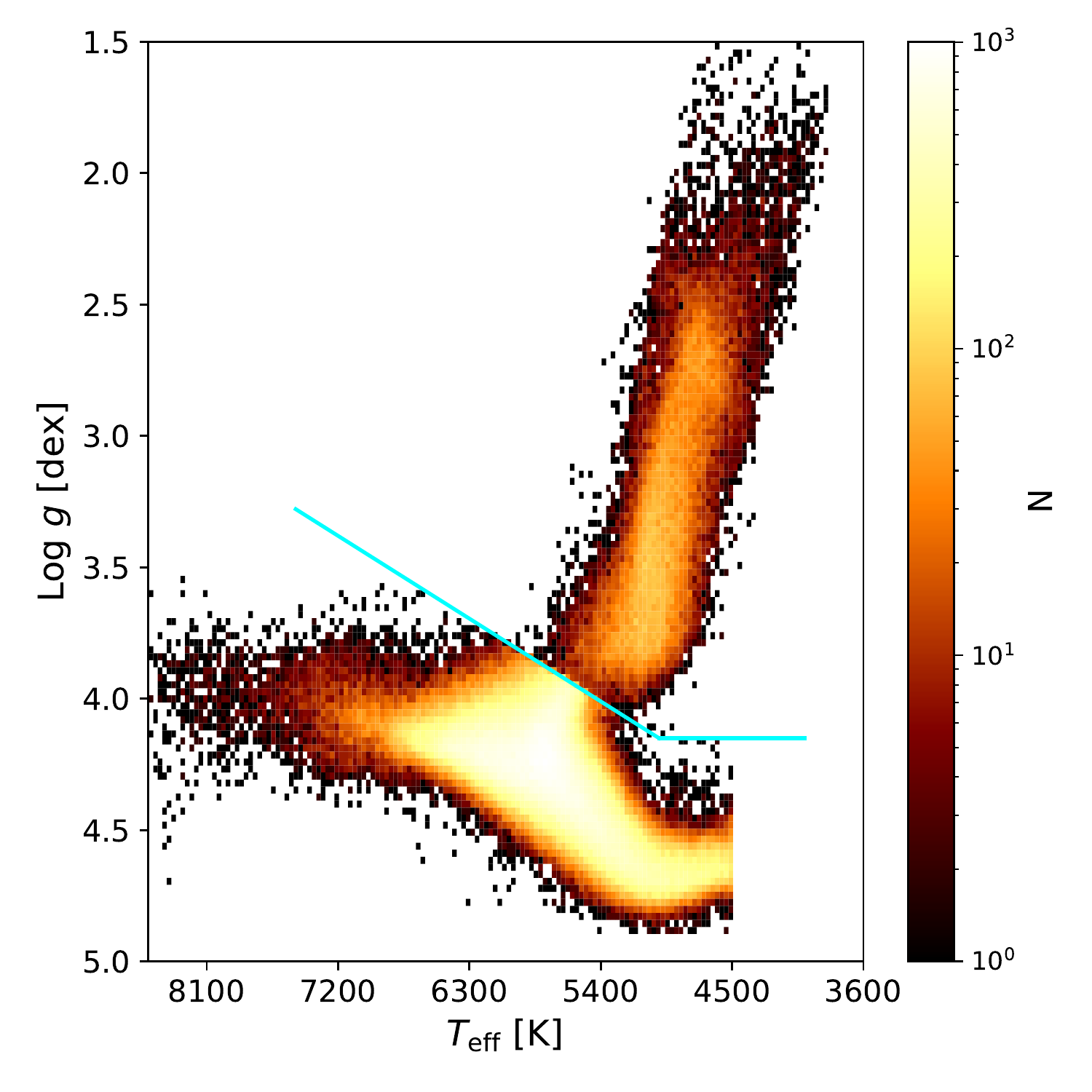} 
    \caption{$T_{\rm eff}$ against log $g$ of sources satisfying the filters described in Sect.2. Below the cyan line are dwarf stars from the LAMOST DR5, while the giant stars are from the value-added RGB catalog of LAMOST DR4.\label{fig:fig1}} 
\end{figure}

Note that all colors referred to hereafter are dereddened intrinsic ones.

Criteria (10) and (11) aim to select high-quality stellar parameters from the LAMOST. For stars with multiple observations, their mean values using inverse variance weighting are adopted.

For criterion (12), given that the \textit{Gaia} satellite scans the sky repeatedly and publishes a mean magnitude for each object, variable stars could be less reliable and should be removed. A number of 112 objects identified as variable stars in the \uppercase{wise} \citep{chen2018}, ATLAs \citep{heinze2018}, ASAS-SN \citep{jaya2020}, Catalina \citep{drake2017}, and ZTF \citep{chen2020} surveys are removed in this step.

Finally, we divide the above sample into dwarfs and giants, considering that they probably suffer different systematic errors in the effective temperature and metallicity. Dwarf stars are selected from $T_{\rm eff}$ against log $g$ diagram as those below the cyan line in Figure \ref{fig:fig1}. For giants, only Red Giant Brunch (RGB) stars are selected (with red clump stars removed) by cross-matching the above sample with a value-added RGB catalog of \uppercase{Lamost} DR4 \citep{wu2019}. The final sample contains 433,845 dwarfs and 55,395 RGB stars, as displayed in Figure \ref{fig:fig1}.

\section{Method and Result} \label{sec:methodresult}

\begin{figure*}[htbp] 
    \centering 
    \includegraphics[width=\textwidth]{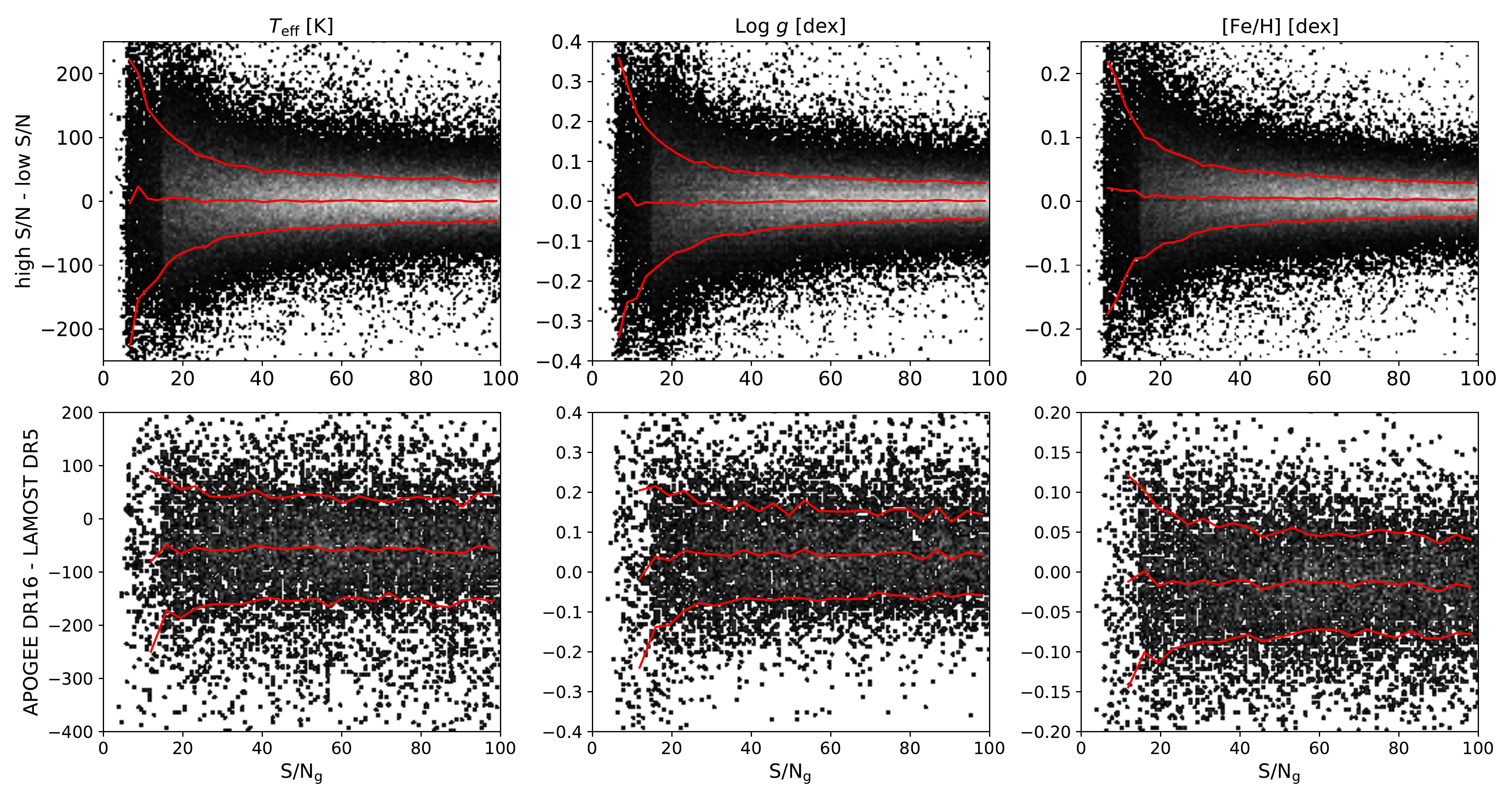} 
    \caption{\textit{Top}: Differences of LAMOST DR5 stellar parameters deduced from two observations of very different $S/N_{\rm g}$ as a function of the lower $S/N_{\rm g}$. The spectrum of higher quality has $S/N_{\rm g} > 100$. The red lines indicate the medians and standard deviations. Note the very small standard deviation values at $S/N_{\rm g} > 30$. 
    \textit{Bottom}: Differences between LAMOST DR5 and APOGEE DR16 stellar parameters for solar-type stars as a function of the LAMOST $S/N_{\rm g}$. The APOGEE stars have spectral S/Ns higher than 100. \label{fig:fig3snr}} 
\end{figure*}

The key idea of the SCR method is that stars of the same stellar atmospheric parameters should have the same intrinsic colors, therefore, plenty of stars with precisely determined spectroscopic parameters can serve as excellent color standards to carry out very precise color calibrations. \citet{yuan2015} demonstrates the method with the SDSS Stripe 82 data. In their work, a well-calibrated control sample within a specific area is selected as a reference to define the intrinsic colors as a function of the stellar parameters, which in turn are used to map out the spatial variations of the color zero-point offset as functions of R.A. and Dec., achieving a remarkable improvement in color calibration by a factor of two to three. Very recently, \citet{huang2020} apply the SCR method to the second data release of the SkyMapper Southern Survey \citep{skymapper}. A uniform calibration with precision better than 1$\%$ is achieved.

In this work, taking advantages of about half million common stars selected in the previous section, we apply this method to the \textit{Gaia} DR2 and obtain the empirical relations between the \textit{Gaia} colors ($G-G_{\rm RP}$ and $G_{\rm BP}-G_{\rm RP}$) and the LAMOST DR5 stellar astrophysical parameters ($T_{\rm eff}$, [Fe/H], log $g$). Since a systematic effect with the $G$ magnitude is known, a control sample within a narrow magnitude range is needed as the reference to explore the magnitude-dependent variations. \textit{Gaia} modifies its instrument configurations (Gates and Windows) according to the $G$ magnitude, even though an iterative calibration is performed to establish the internal photometric system, the convergence problems still exist among the different configurations, and stars around these joints may have relatively larger uncertainties \citep{evans2018}. Considering the distribution of the Gates and Window Classes \citep{evans2018} as well as the number of stars, stars of $13.3 < G < 13.7$ are selected as the control sample, consisting of 50,343 dwarfs and 6,144 RGB stars. 

\begin{figure*}[htbp] 
    \centering 
    \includegraphics[width=\textwidth]{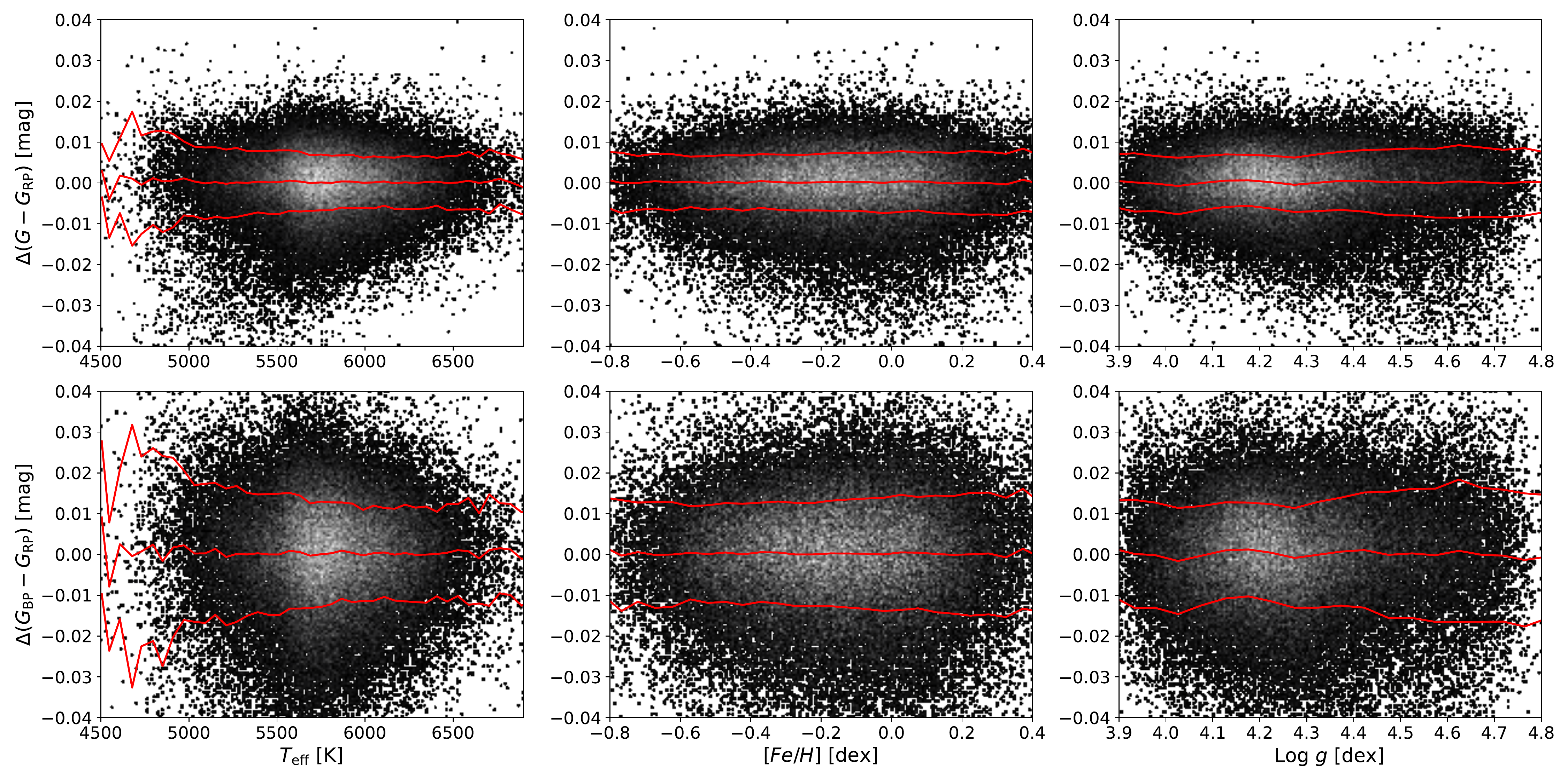} 
    \caption{Fitting residuals of the $G-G_{\rm RP}$ (\textit{top}) and $G_{\rm BP}-G_{\rm RP}$ (\textit{bottom}) of the control sample for the dwarf stars as functions of $T_{\rm eff}$, [Fe/H], and log $g$. The red lines show the medians and standard deviations of the residuals.\label{fig:fig2ms}} 
\end{figure*}

\begin{figure*}[htbp] 
    \centering 
    \includegraphics[width=\textwidth]{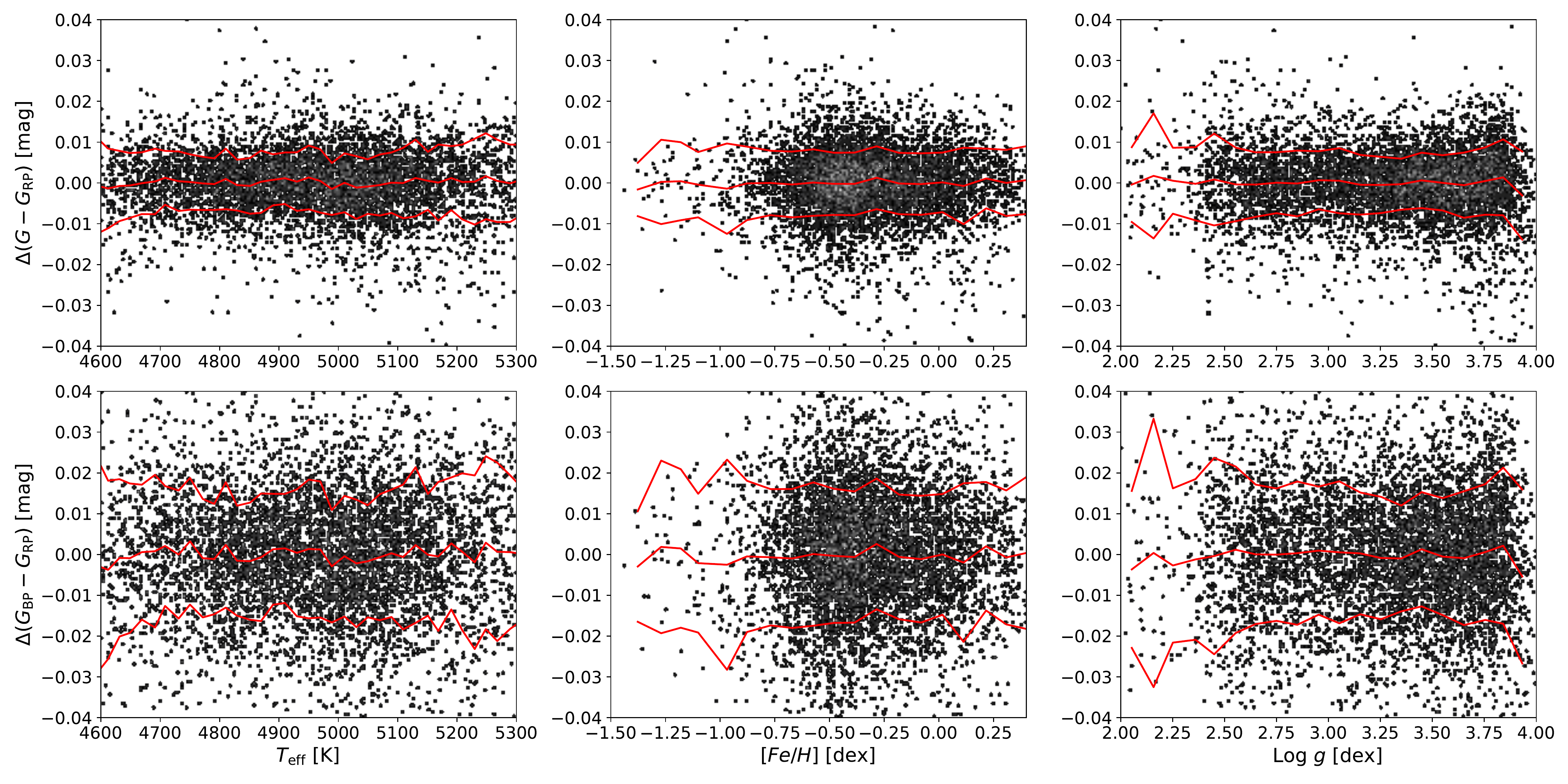} 
    \caption{Same as Figure \ref{fig:fig2ms}, but for the RGB stars.\label{fig:fig2rgb}} 
\end{figure*}

Magnitude-dependent systematic effects in the LAMOST stellar parameters could cause false trends with $G$ magnitudes in our results. We select multiply-observed stars in the LAMOST DR5 to investigate such effects and find that there are no systematic effects with the $S/N_{\rm g}$ till at very low $S/N_{\rm g}$ of about 15. The results are plotted in the top panels of Figure \ref{fig:fig3snr} showing a comparison between observations of very different $S/N_{\rm g}$. To further verify this result, we cross-match the LAMOST DR5 with the APOGEE DR16 \citep{apogee16} and select a sample of solar-type stars to plot their stellar parameter differences as a function of LAMOST $S/N_{\rm g}$ in the bottom panels of Figure \ref{fig:fig3snr}. A flat offset is seen for each stellar parameter at $S/N_{\rm g} > 15$, consistent with the test result from duplicated stars. Note the offsets are not exactly zero due to systematic errors in the LAMOST and APOGEE stellar parameters.

To precisely construct the relationships between the \textit{Gaia} colors and the LAMOST stellar parameters, we divide our control sample into different bins as following: $T_{\rm eff}$ in 300 K, [Fe/H] in 0.2 dex, and log $g$ in 0.3 dex for dwarfs; $T_{\rm eff}$ in 300 K, [Fe/H] in 0.5 dex, and log $g$ in 0.4 dex for RGB stars. A linear fitting formula is adopted for each grid: $C=a_0+a_1 \times T_{\rm eff}+a_2 \times [Fe/H]+a_3 \times$ log $g$, where C can be ($G_{\rm BP}-G_{\rm RP}$)$_0$ or ($G-G_{\rm RP}$)$_0$. 2$\sigma$-clipping is performed during the fitting process. Bins containing less than 50 stars are excluded to avoid poor fitting. The fitting residuals are plotted in Figure \ref{fig:fig2ms} and Figure \ref{fig:fig2rgb}. The median residuals of both dwarfs and RGB giants are very close to zero. The typical standard deviations are about 7.5 mmag and 15 mmag for $G-G_{\rm RP}$ and $G_{\rm BP}-G_{\rm RP}$, respectively.

\begin{figure}[htbp] 
    \centering 
    \includegraphics[width=3.5in]{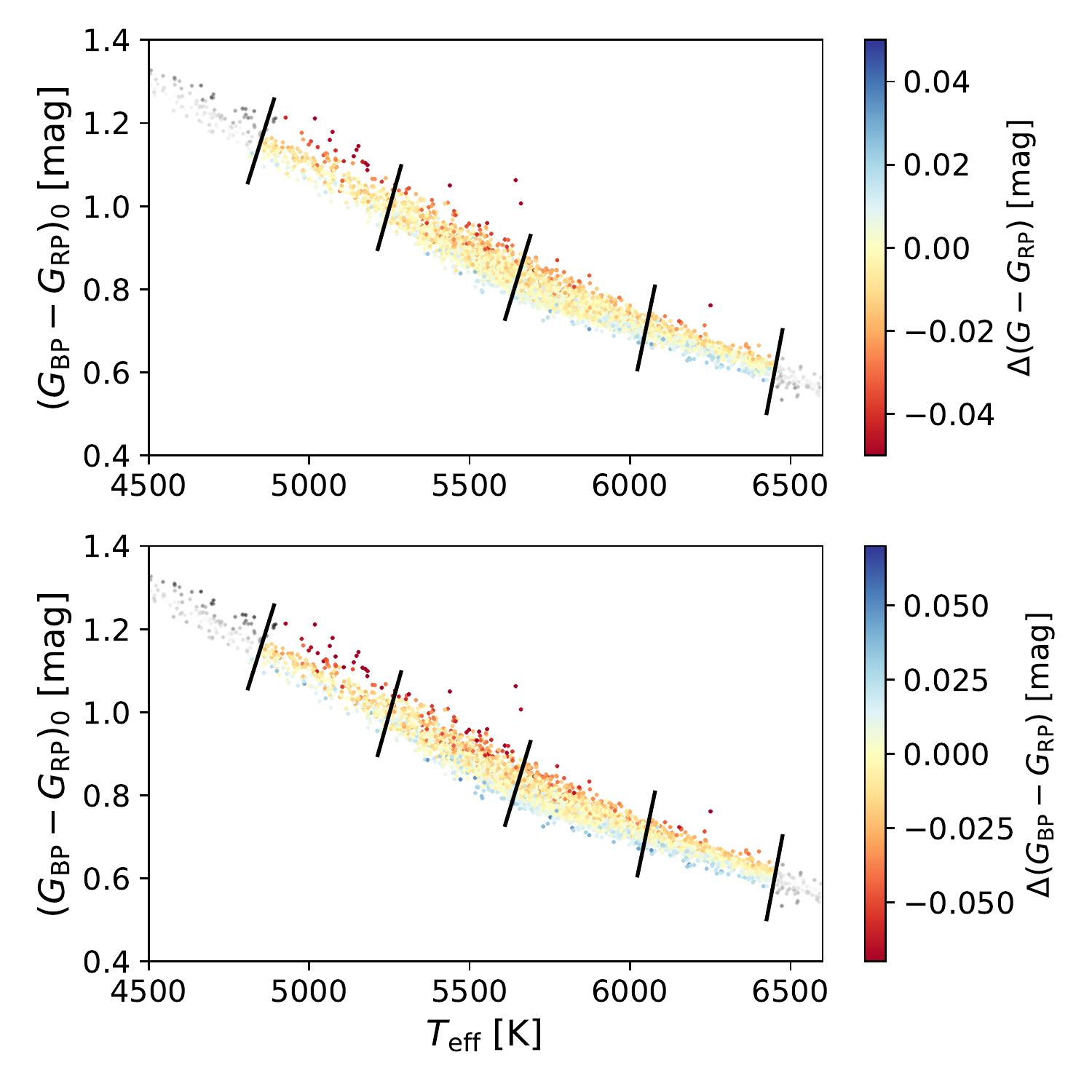} 
    \caption{Distributions of the color residuals (fitted colors $-$ observed colors) in the $T_{\rm eff}$ and ($G_{\rm BP}-G_{\rm RP}$)$_0$ plane for the dwarf stars. The black solid lines mark the boundaries of different subsamples. To avoid the selection effect, only colorful points are used.  Only 1 in 100 stars are plotted. \label{fig:fig3ms}}
\end{figure}

\begin{figure}[htbp] 
    \centering 
    \includegraphics[width=3.5in]{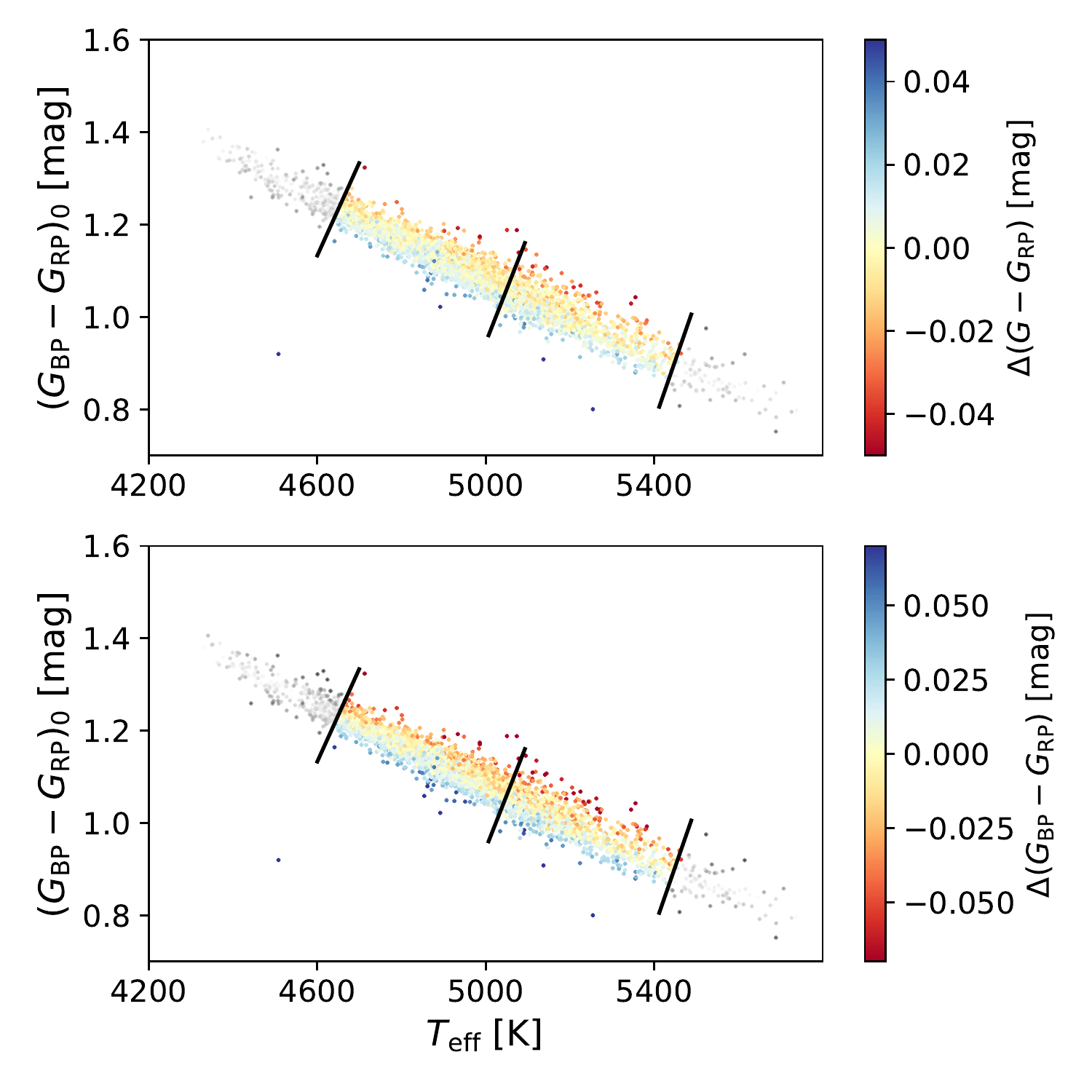} 
    \caption{Same as Figure \ref{fig:fig3ms}, but for the RGB stars. Only 1 in 10 stars are plotted. \label{fig:fig3rgb}} 
\end{figure}

Applying the empirical relations derived above to the whole sample, color residuals/corrections (fitted colors $-$ observed colors) are obtained. Their distributions in the $T_{\rm eff}$ and ($G_{\rm BP}-G_{\rm RP}$)$_0$ plane are shown in Figure \ref{fig:fig3ms} and Figure \ref{fig:fig3rgb}. At a given effective temperature, stars of redder intrinsic colors have typically smaller (negative) residuals. To examine the color dependence, both dwarfs and RGB giants are divided into subsamples by the black solid lines in Figure \ref{fig:fig3ms} and Figure \ref{fig:fig3rgb}. Four subsamples are adopted for dwarfs (Main sequence, hereafter MS), with typical ($G_{\rm BP}-G_{\rm RP}$)$_0$ colors of 1.06, 0.90, 0.77, and 0.66 mag, respectively. Two subsamples are adopted for RGB stars, with typical ($G_{\rm BP}-G_{\rm RP}$)$_0$ colors of 1.13 and 1.01 mag, respectively. The sample numbers from red to blue are 32655, 102441, 163325, 68493 for dwarfs and 24697, 14601 for giants. The gray points are discarded because stars there are partly dropped due to the requirement N $>$ 50, and the dramatic change of numbers at the margin can cause bias of color residuals. 

\begin{figure*}[htbp] 
    \centering 
    \includegraphics[width=\textwidth]{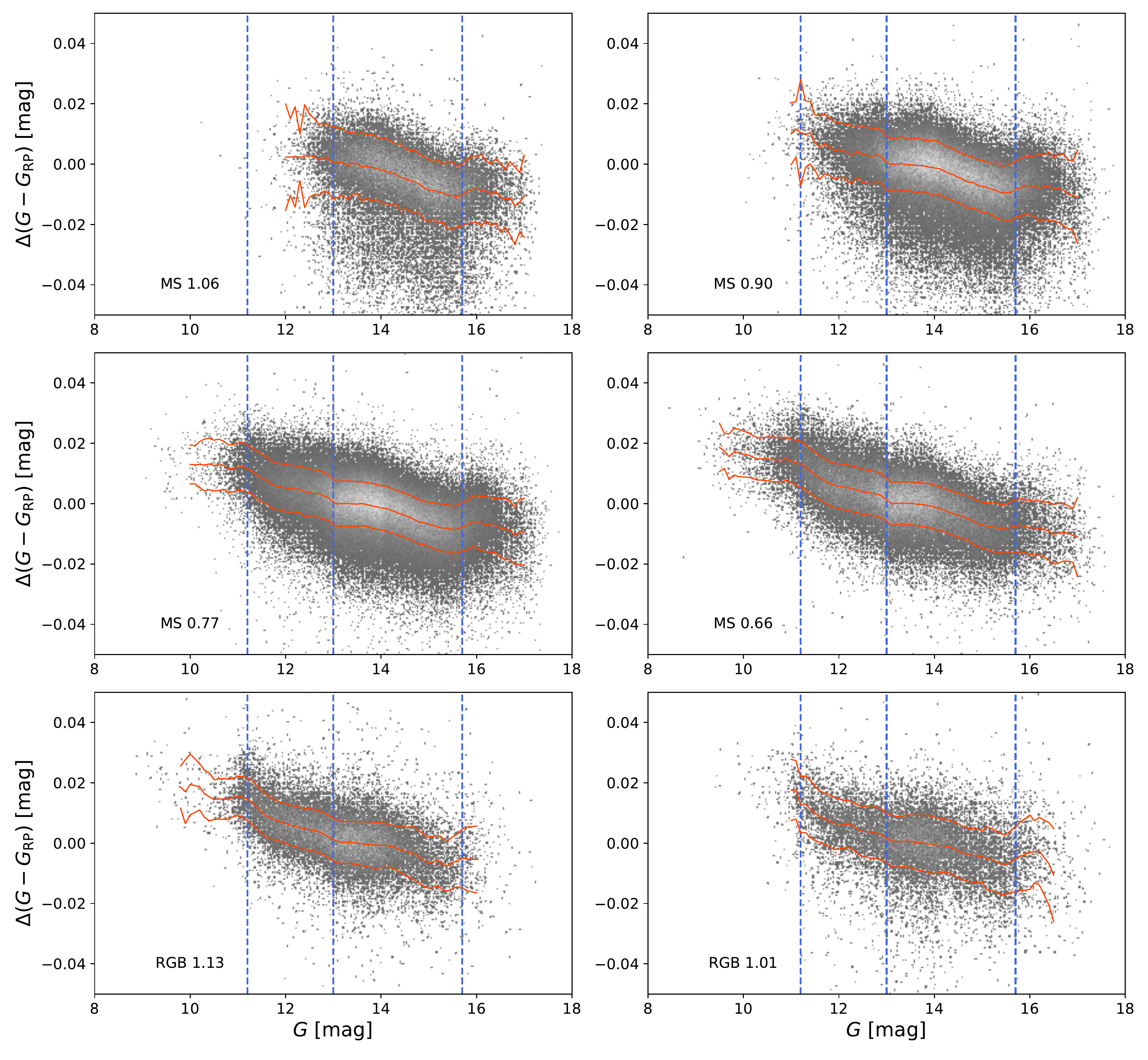} 
    \caption{Fitting residuals of $G-G_{\rm RP}$ against $G$ magnitude for each subsample from Figure \ref{fig:fig3ms} and Figure \ref{fig:fig3rgb} , labeled by its evolution stage and typical ($G_{\rm BP}-G_{\rm RP}$)$_0$ color. Each subsample is binned along the $G$ magnitude with 0.05 mag interval. The median and standard deviation values are estimated and shown by the orange lines. A 2$\sigma$-clipping is performed in this process. The changes of instrument configurations are marked by the vertical blue lines. \label{fig:fig4G-RP} } 
\end{figure*}

\begin{figure*}[t!] 
    \centering 
    \includegraphics[width=\textwidth]{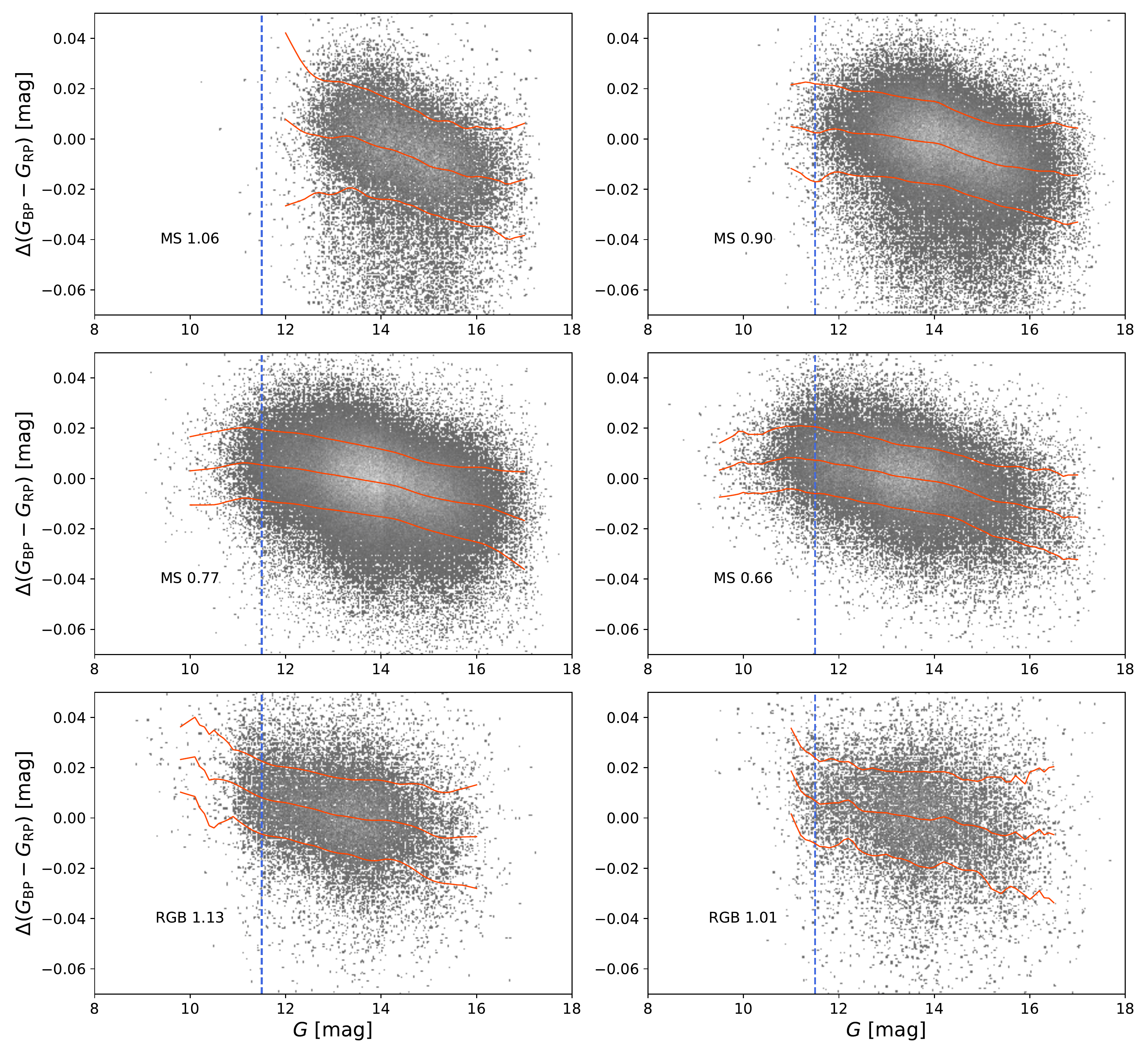} 
    \caption{Same as Figure \ref{fig:fig4BP-RP}, but for the $G_{\rm BP}-G_{\rm RP}$ color.\label{fig:fig4BP-RP}} 
\end{figure*}

For each subsample, the distributions of its color residuals against $G$ magnitudes are plotted in Figure \ref{fig:fig4G-RP} and Figure \ref{fig:fig4BP-RP}. The points are grouped into bins along the $G$ magnitude for every 0.05 mag. The median and standard deviation values are over-plotted in orange lines. Typical standard deviations are $\sim$ 7 -- 9 mmag for $G-G_{\rm RP}$ and $\sim$ 12 -- 17 mmag for $G_{\rm BP}-G_{\rm RP}$. Magnitude dependent offsets are clearly seen for all subsamples. The offsets (median of the color residuals) are smoothed by the locally estimated scatterplot smoothing (LOWESS) and taken as the calibration curves. The algorithm works by taking the $frac \times N$ closest points to each data point and estimating smoothed $y$ values using a weighted linear regression based on their $x$ distances, where $N$ is the total number of points and $frac$ varies for subsamples with typically 0.05. Note that the curves of different subsamples may have different magnitude ranges. In places with few sources at the very bright or faint ends, linear extrapolation is performed. Specifically, corrections can be expressed as:
\begin{equation}\label{eq3}
    C^{'} = C + \Delta C
\end{equation}
where $C^{'}$ is the corrected color, $C$ is the \textit{Gaia} DR2 color and $\Delta C$ is the color correction term. 

In Figure \ref{fig:fig4G-RP}, there are three marked variations at around $G$ $\sim$ 11.2, 13.0, and 15.7 mag that are probably related to the changes of instrument configurations. Gates are implemented on the \textit{Gaia} CCDs to ease the saturation issues in the case of bright stars, e.g. $G$ $<$ 12 mag. Sizes of Windows in along-scan and across-scan direction are modified at $G$ = 13.0 and 16 mag for the $G$ band, and at $G$ = 11.5 for the $G_{\rm BP}$ and $G_{\rm RP}$ bands \citep{evans2018}. However, the choice of the instrument configurations depends on the observed signals in the $G$ band, the corresponding calibrated $G$ magnitudes are not exactly identical for every single observation. For example, the transition of the $G$ band from Window Class 0 to 1 may move around $G$ = 13 mag, and the transition of the $G_{\rm BP}$ and $G_{\rm RP}$ may move around $G$ = 11.5 mag, due to issues including flat-fielding and stray light contamination. Since the homogeneous calibration adopted by \textit{Gaia} is a function of the $G$ magnitude and may differ significantly before and after the transition magnitude, some corrections could be incorrect. Consequently, the random uncertainties around the transitions are larger than those in adjacent regions, as demonstrated by Figure 9 in \citet{evans2018}. The features at $G$ $\sim$ 11.2, 13.0, and 15.7 mag in Figure \ref{fig:fig4G-RP} are consistent with Figure 9 in \citet{evans2018}, suggesting that these fine structures are real and accurate. 

\begin{figure}[htbp] 
    \centering 
    \includegraphics[width=3.5in]{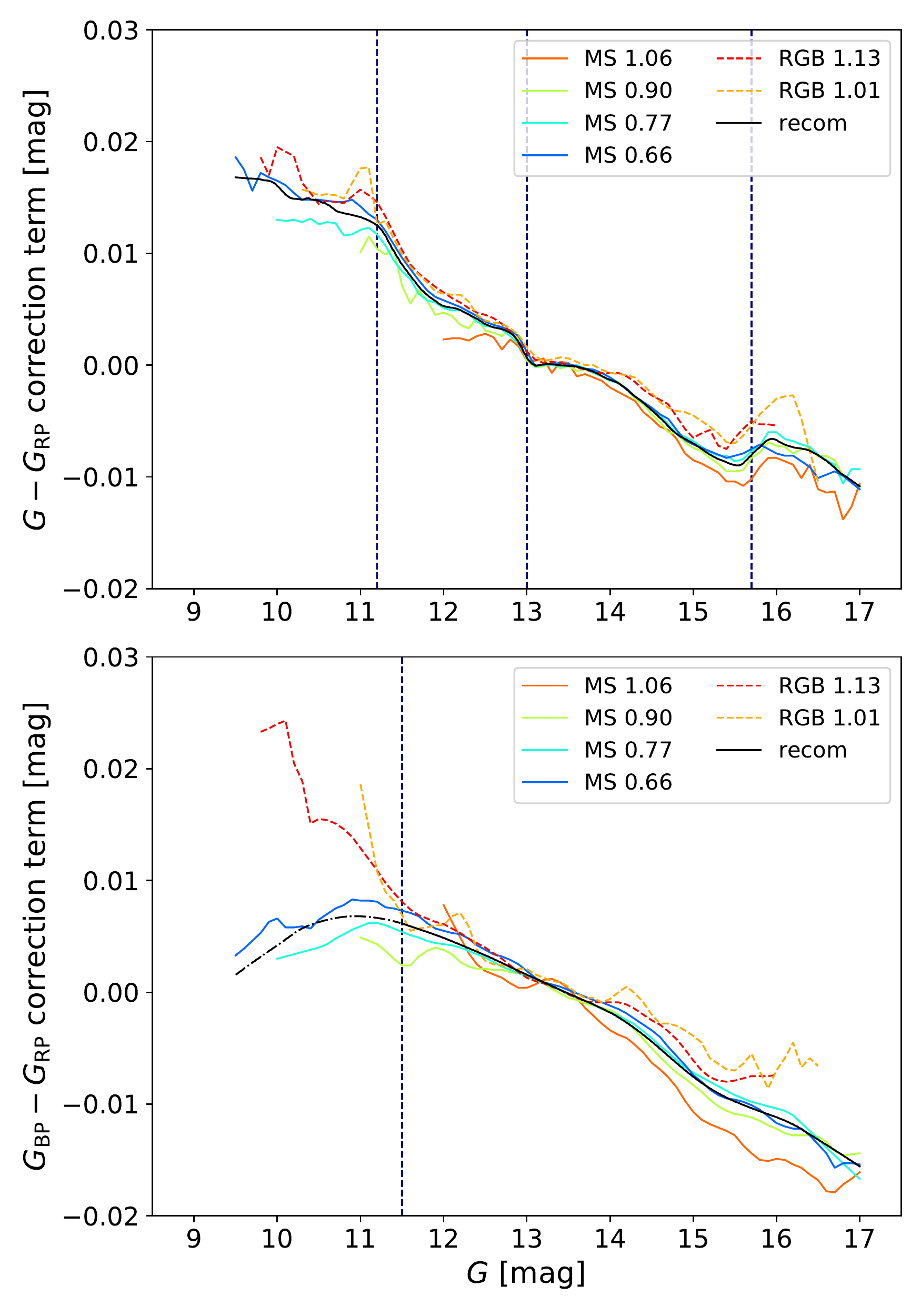} 
    \caption{Color-dependent and recommended calibration curves for $G-G_{\rm RP}$ and $G_{\rm BP}-G_{\rm RP}$ . \label{fig:fig5}} 
\end{figure}

The final results and a comparison of calibrations of different subsamples are plotted in Figure \ref{fig:fig5}. Positions of changes of Gates and Windows are marked as well. The analysis and possible applications of our results are discussed in Section \ref{sec:dis}. 

\begin{figure}[htbp] 
    \centering 
    \includegraphics[width=3.5in]{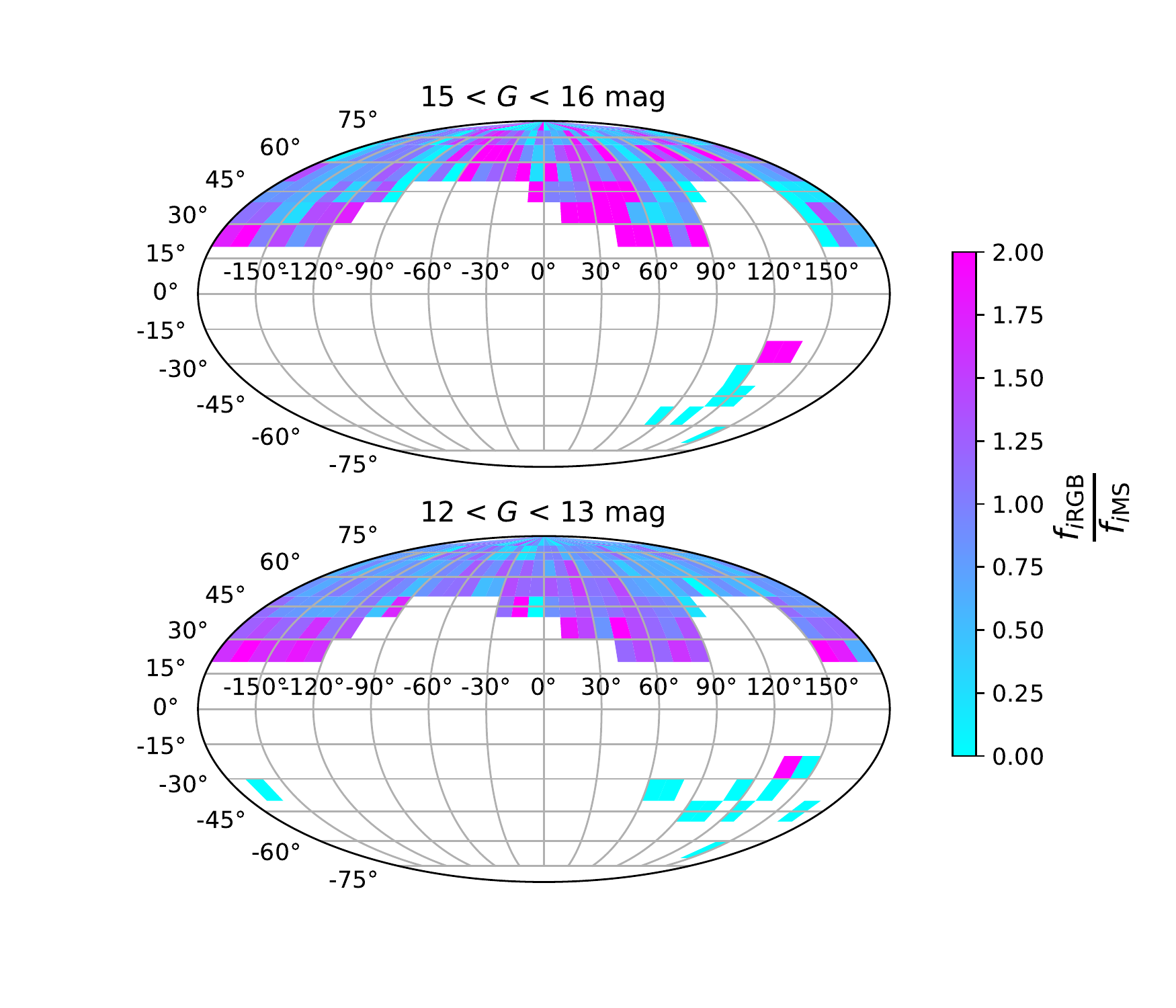} 
    \caption{Comparisons of spatial distributions in Mollweide projection of the dwarfs and RGB giants of 12 $< G <$ 13 mag (\textit{top}) and 15 $< G <$ 16 mag (\textit{bottom}). 
    \label{fig:figred}} 
\end{figure}

\section{Discussion} \label{sec:dis}

\subsection{Color dependence}

As shown in Figure \ref{fig:fig5}, despite places with few sources and relatively larger errors at the very bright ends, the calibration curves of $G-G_{\rm RP}$ yielded by the six different subsamples agree well with each other when $G$ $<$ 14 mag. However, when $G$ $>$ 14 mag, discrepancies exist between the curves of the two RGB subsamples (RGB 1.13 and RGB 1.01) and the red dwarf subsample (MS 1.06), while those of the other three blue dwarf subsamples (MS 0.90, MS 0.77, and MS 0.66) still agree well. As for the calibration curves for $G_{\rm BP}-G_{\rm RP}$, the same trend exists as well when $G$ $>$ 14 mag.  When $G$ $<$ 11.5 mag, the curves are divided into two groups (RGB 1.13 and RGB 1.01 versus MS 0.77 and MS 0.66) according to their colors. 

The discrepancies of the three red lines (MS 1.06, RGB 1.13, and RGB 1.01) at $G$ $>$ 14 mag in Figure \ref{fig:fig5} are not real, but caused by the combination of two reasons: the selection function of the LAMOST data and the spatially dependent systematics of the SFD reddening map. The LAMOST surveys have four different types of plates: very bright (VB), bright (B), median bright (M), and faint (F) plates, in which VB plates target stars of $r <$ 14 mag while the others target fainter stars \citep{chen2018lamost}. Compared to blue stars, bright red stars are more likely to be giants while faint red stars to be dwarfs. The comparisons of the spatial distributions of dwarfs and RGB stars are shown in Figure \ref{fig:figred}. The whole sky is divided into different bins of 10 degree by 10 degree in size. The percentages $f_i=\frac{N_i}{N_{\rm total}}$ of RGB stars and dwarfs in each bin are calculated ,and the colorbars represent the ratios of dwarf percentage to RGB percentage $\frac{f_{i \rm RGB}}{f_{i \rm MS}}$. The closer the ratios get to 1, the smaller the differences between the spatial distributions of RGB stars and dwarfs are. Clearly, the differences are larger when $G >$ 14 mag. Note that the \textit{Gaia} $G$ magnitudes are very close to the SDSS $r$ magnitudes for typical stars.

\begin{figure}[htbp] 
    \centering 
    \includegraphics[width=3.5in]{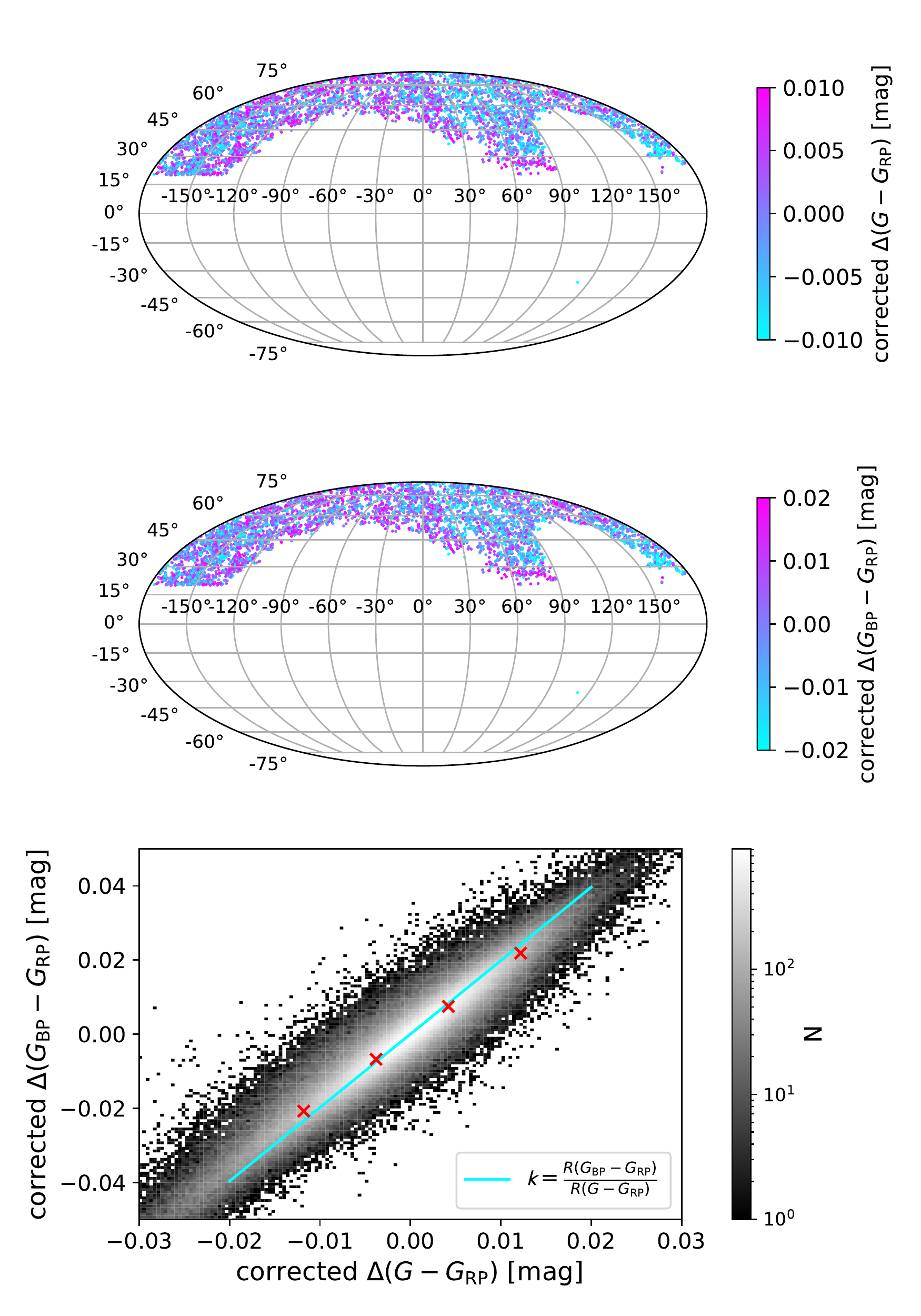} 
    \caption{Bias of the SFD dust reddening map. \textit{Top}: Scatter plot of the sky distribution in Mollweide projection of the corrected $G-G_{\rm RP}$ residual (fitted colors $-$ observed colors). \textit{Middle}: Same but for $G_{\rm BP}-G_{\rm RP}$. Only 10000 randomly selected points are plotted. \textit{Bottom}: Comparison of the corrected color residuals. The red crosses are the median values of two corrected residuals. Instead of fitting the red crosses, the slope of the cyan line is determined by the median value of the $\frac{R(G_{\rm BP}-G_{\rm RP})}{R(G-G_{\rm RP})}$ ratios of the sample stars. \label{fig:fig67}} 
\end{figure}

On the other hand, the SFD reddening map is found to show systematic errors that depend on spatial position and dust temperature (e.g., \citealp{peek2010}; Sun et al., to be submitted). The top and middle panels of Figure \ref{fig:fig67} show sky distributions of $G-G_{\rm RP}$ and $G_{\rm BP}-G_{\rm RP}$ residuals (fitted colors $-$ observed colors) after correcting for the trends with $G$ magnitude, which are almost identical and therefore caused by the same problem. The SFD map overestimates reddening in the pink regions (positive residuals) while underestimates reddening in the blue regions (negative residuals), which is consistent with the result of Sun et al. (to be submitted) well. A strong correlation is seen in the the bottom panel of Figure \ref{fig:fig67} that compares the corrected $G-G_{\rm RP}$ and $G_{\rm BP}-G_{\rm RP}$ residuals of the whole sample. The red crosses are the median values of two corrected residuals. They agree very well with the cyan line, whose slope is determined by the median value ($\sim$ 1.99) of the $\frac{R(G_{\rm BP}-G_{\rm RP})}{R(G-G_{\rm RP})}$ ratios of the sample, as expected. 

For the $G_{\rm BP}$ and $G_{\rm RP}$ bands, even though the \textit{Gaia} photometric pipeline treats both 2D and 1D Windows as aperture photometry in the same way \citep{evans2018}, different instrument configurations may still cause problems. \citetalias{wei2018} points out that convergence during the calibration process may be dominated by stars having the most number, which can cause discrepancies among colors. They find that $G_{\rm BP}$ magnitude has a systematic tendency among $G_{\rm BP}-G_{\rm RP}$ and their offset happens around 10.99 mag. \citetalias{maw2018} further confirms that the color-dependent break happens in $G_{\rm BP}$ band at $G$ $\sim$ 10.87 mag, by analyzing the uncertainties of the $G_{\rm BP}$ fluxes. Likewise, we believe that the color-dependent feature at $G$ $<$ 11.5 mag in our result of $G_{\rm BP}-G_{\rm RP}$ calibration is due to the same reason.

We combine the three unbiased dwarf subsamples (MS 0.90, MS 0.77, and MS 0.66) together to determine the recommended calibration curves with the same procedure in Section \ref{sec:methodresult}. The recommended calibration curves are overplotted in black lines in Figure \ref{fig:fig5}. The differences between the recommended curve and the three colored ones (MS 0.90, MS 0.77, and MS 0.66) are computed, with standard deviations 0.3 and 0.6 mmag for $G-G_{\rm RP}$ and $G_{\rm BP}-G_{\rm RP}$, respectively. This suggests that the calibration curve of each subsample has a typical random error smaller than 1.0 mmag. Note that the recommended curves work well in most cases except for $G_{\rm BP}-G_{\rm RP}$ when $G$ $<$ 11.5 mag, which is plotted in the dash-dot line. The recommended color calibration curves as well as the calibration curves of different subsamples are listed in Table \ref{tab:1}.

\begin{table*}[t]
\centering
     \caption{Color calibration curves. The first column is $G$ magnitude. The second and third columns are respectively recommended $G-G_{\rm RP}$ and $G_{\rm BP}-G_{\rm RP}$ calibration curves, followed by those yielded by different subsamples. All calibration curves are in units of mmag. \label{tab:1}}
    \begin{longtable}{c|cc|cc|cc|cc|cc|cc|cc} 
     \textbf{G} & \multicolumn{2}{c|}{\textbf{recom}} & \multicolumn{2}{c|}{\textbf{MS 1.06}} & \multicolumn{2}{c|}{\textbf{MS 0.89}} & \multicolumn{2}{c|}{\textbf{MS 0.77}} & \multicolumn{2}{c|}{\textbf{MS 0.67}} & \multicolumn{2}{c|}{\textbf{RGB 1.13}} & \multicolumn{2}{c}{\textbf{RGB 1.02}} \\
     \hline
... & ... & ... & ... & ... & ... & ... & ... & ... & ... & ... & ... & ... & ... & ... \\
     
 13.01 & 0.57 & 1.53 & 0.40 & 0.43 & 0.34 & 1.37 & 0.53 & 1.46 & 1.07 & 1.84 & 1.13 & 1.27 & 1.52 & 2.05 \\ 
13.02 & 0.46 & 1.50 & 0.40 & 0.46 & 0.28 & 1.34 & 0.46 & 1.42 & 0.94 & 1.78 & 1.06 & 1.24 & 1.44 & 2.00 \\ 
13.03 & 0.37 & 1.46 & 0.40 & 0.49 & 0.22 & 1.31 & 0.39 & 1.38 & 0.81 & 1.72 & 0.99 & 1.21 & 1.36 & 1.95 \\ 
13.04 & 0.28 & 1.43 & 0.40 & 0.52 & 0.16 & 1.28 & 0.32 & 1.34 & 0.68 & 1.66 & 0.92 & 1.18 & 1.28 & 1.90 \\ 
13.05 & 0.21 & 1.39 & 0.40 & 0.55 & 0.10 & 1.25 & 0.25 & 1.30 & 0.55 & 1.60 & 0.85 & 1.15 & 1.20 & 1.85 \\ 
13.06 & 0.14 & 1.36 & 0.40 & 0.58 & 0.04 & 1.22 & 0.18 & 1.26 & 0.42 & 1.54 & 0.78 & 1.12 & 1.12 & 1.80 \\ 
13.07 & 0.09 & 1.33 & 0.40 & 0.61 & -0.02 & 1.19 & 0.11 & 1.22 & 0.29 & 1.48 & 0.71 & 1.09 & 1.04 & 1.75 \\ 
13.08 & 0.05 & 1.29 & 0.40 & 0.64 & -0.08 & 1.16 & 0.04 & 1.18 & 0.16 & 1.42 & 0.64 & 1.06 & 0.96 & 1.70 \\ 
13.09 & 0.01 & 1.26 & 0.40 & 0.67 & -0.14 & 1.13 & -0.03 & 1.14 & 0.03 & 1.36 & 0.57 & 1.03 & 0.88 & 1.65 \\ 
13.10 & -0.01 & 1.23 & 0.40 & 0.70 & -0.20 & 1.10 & -0.10 & 1.10 & -0.10 & 1.30 & 0.50 & 1.00 & 0.80 & 1.60 \\ 
13.11 & -0.02 & 1.19 & 0.42 & 0.74 & -0.19 & 1.07 & -0.09 & 1.07 & -0.08 & 1.26 & 0.47 & 0.98 & 0.76 & 1.57 \\ 
13.12 & -0.02 & 1.16 & 0.44 & 0.78 & -0.18 & 1.04 & -0.08 & 1.04 & -0.06 & 1.22 & 0.44 & 0.96 & 0.72 & 1.54 \\ 
13.13 & -0.02 & 1.13 & 0.46 & 0.82 & -0.17 & 1.01 & -0.07 & 1.01 & -0.04 & 1.18 & 0.41 & 0.94 & 0.68 & 1.51 \\ 
13.14 & -0.01 & 1.09 & 0.48 & 0.86 & -0.16 & 0.98 & -0.06 & 0.98 & -0.02 & 1.14 & 0.38 & 0.92 & 0.64 & 1.48 \\ 
13.15 & 0.00 & 1.06 & 0.50 & 0.90 & -0.15 & 0.95 & -0.05 & 0.95 & -0.00 & 1.10 & 0.35 & 0.90 & 0.60 & 1.45 \\ 
13.16 & 0.02 & 1.03 & 0.52 & 0.94 & -0.14 & 0.92 & -0.04 & 0.92 & 0.02 & 1.06 & 0.32 & 0.88 & 0.56 & 1.42 \\ 
13.17 & 0.03 & 1.00 & 0.54 & 0.98 & -0.13 & 0.89 & -0.03 & 0.89 & 0.04 & 1.02 & 0.29 & 0.86 & 0.52 & 1.39 \\ 
13.18 & 0.04 & 0.96 & 0.56 & 1.02 & -0.12 & 0.86 & -0.02 & 0.86 & 0.06 & 0.98 & 0.26 & 0.84 & 0.48 & 1.36 \\ 
13.19 & 0.05 & 0.93 & 0.58 & 1.06 & -0.11 & 0.83 & -0.01 & 0.83 & 0.08 & 0.94 & 0.23 & 0.82 & 0.44 & 1.33 \\ 
13.20 & 0.06 & 0.90 & 0.60 & 1.10 & -0.10 & 0.80 & -0.00 & 0.80 & 0.10 & 0.90 & 0.20 & 0.80 & 0.40 & 1.30 \\ 

... & ... & ... & ... & ... & ... & ... & ... & ... & ... & ... & ... & ... & ... & ... \\
    \end{longtable}
\tablecomments{Table 1 is published in its entirety in the machine-readable format. A portion is shown here for guidance regarding its form and content.}
\end{table*}

\subsection{Comparisons with previous works}

\citet{evans2018} publish a set of revised sensitivity curves called REV, which is preferred but have no significant differences with the DR2 ones. In \citetalias{casa2018}, with the well-calibrated CALSPEC spectral library, they compute the synthetic $G$, $G_{\rm BP}$, and $G_{\rm RP}$ magnitudes using the REV curves and compare with those in the \textit{Gaia} DR2 catalog, yielding a linear correction of 3.5 mmag/mag in the $G$ magnitude. \citetalias{wei2018} reconstructs the \textit{Gaia} passbands in basic functional analytic framework with four spectral libraries including the CALSPEC and finds a close magnitude drift with \citetalias{casa2018}. Later, \citetalias{maw2018} updates the spectral libraries and applies the same method in \citetalias{wei2018}, suggesting a 3.2 mmag/mag correction in the $G$ magnitude.

\begin{figure}[htbp] 
    \centering 
    \includegraphics[width=3.5in]{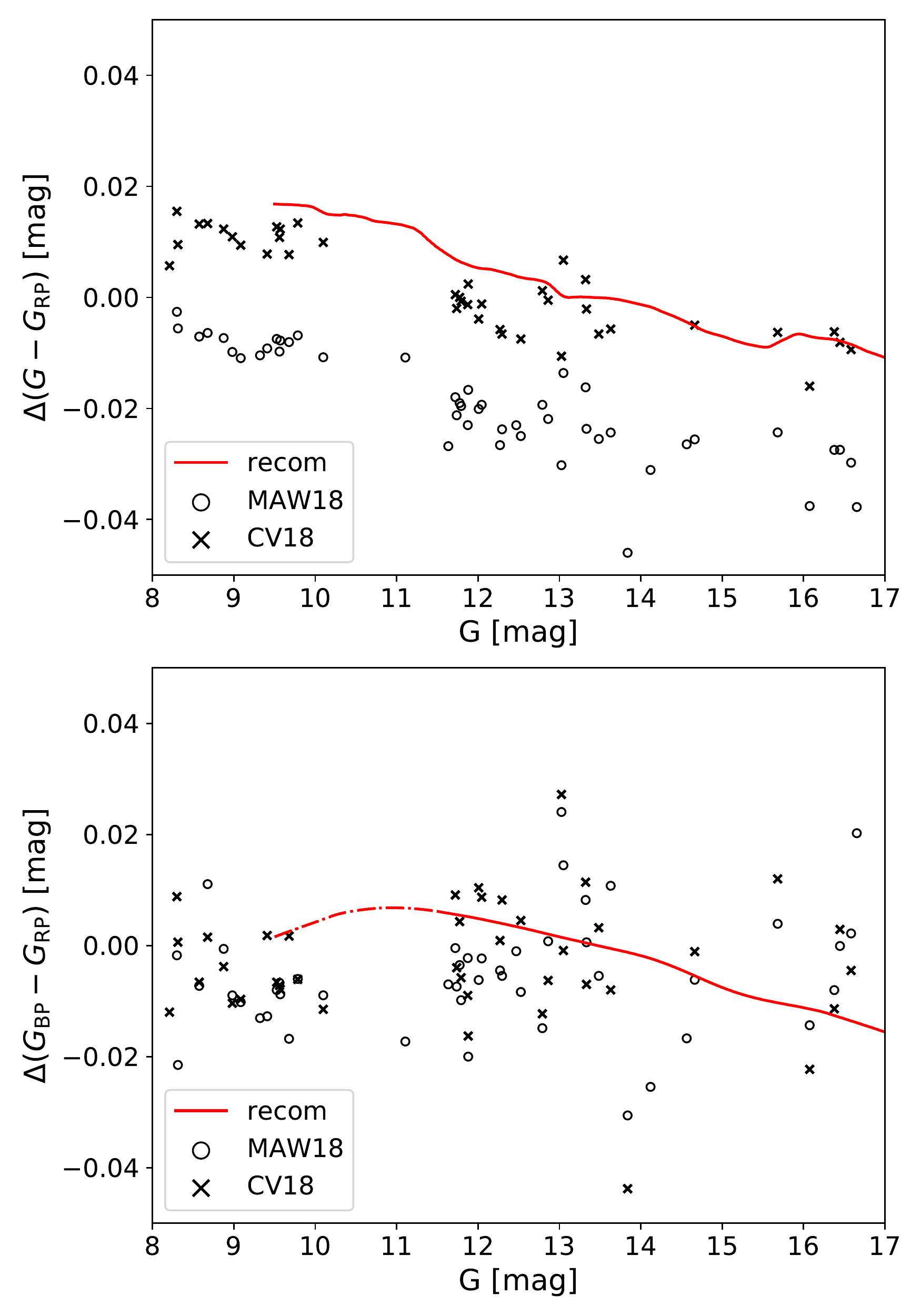}  
    \caption{Comparisons of our results (red lines) with those of \citetalias{maw2018} (empty circles) and \citetalias{casa2018} (black crosses). Note that the typical errors are about 10 mmag for the \citetalias{maw2018} and \citetalias{casa2018} data.\label{fig:fig13}} 
\end{figure}

Figure \ref{fig:fig13} compares our results with those of \citetalias{casa2018} and \citetalias{maw2018}. For the $G-G_{\rm RP}$ color, systematic offsets can be seen between any two of the three references because of the different processing progress: (1) we take $G$ = 13.5 $\pm$ 0.2 mag as the control sample; (2) \citetalias{casa2018} applies the REV transmission curves and the official zero-points; and (3) \citetalias{maw2018} uses their own transmission curves called MAW and a zero-point of 0. Both \citetalias{casa2018} and \citetalias{maw2018} provide a linear fitting of the systematic trend in the $G$ band. Our result shows a very similar trend, consistent with the results of \citetalias{casa2018} and \citetalias{maw2018}. However, thanks to a much larger sample of stars used (half million compared to about one hundred), more fine details are revealed in this work, particularly those due to the changes of the instrument configurations as discussed in Section \ref{sec:methodresult}. For the $G_{\rm BP}-G_{\rm RP}$ color, our result is also consistent with those of \citetalias{casa2018} and \citetalias{maw2018} within their errors. Our result shows a small but clear trend with G magnitude at G $>$ 11.5 mag. At the bright end, our result is slightly higher, probably due to the fact that most stars there in \citetalias{casa2018} and \citetalias{maw2018} are bluer.

\subsection{Revised color-color diagram}\label{dis_colorcolor}

\begin{figure*}[htbp] 
    \centering 
    \includegraphics[width=\textwidth]{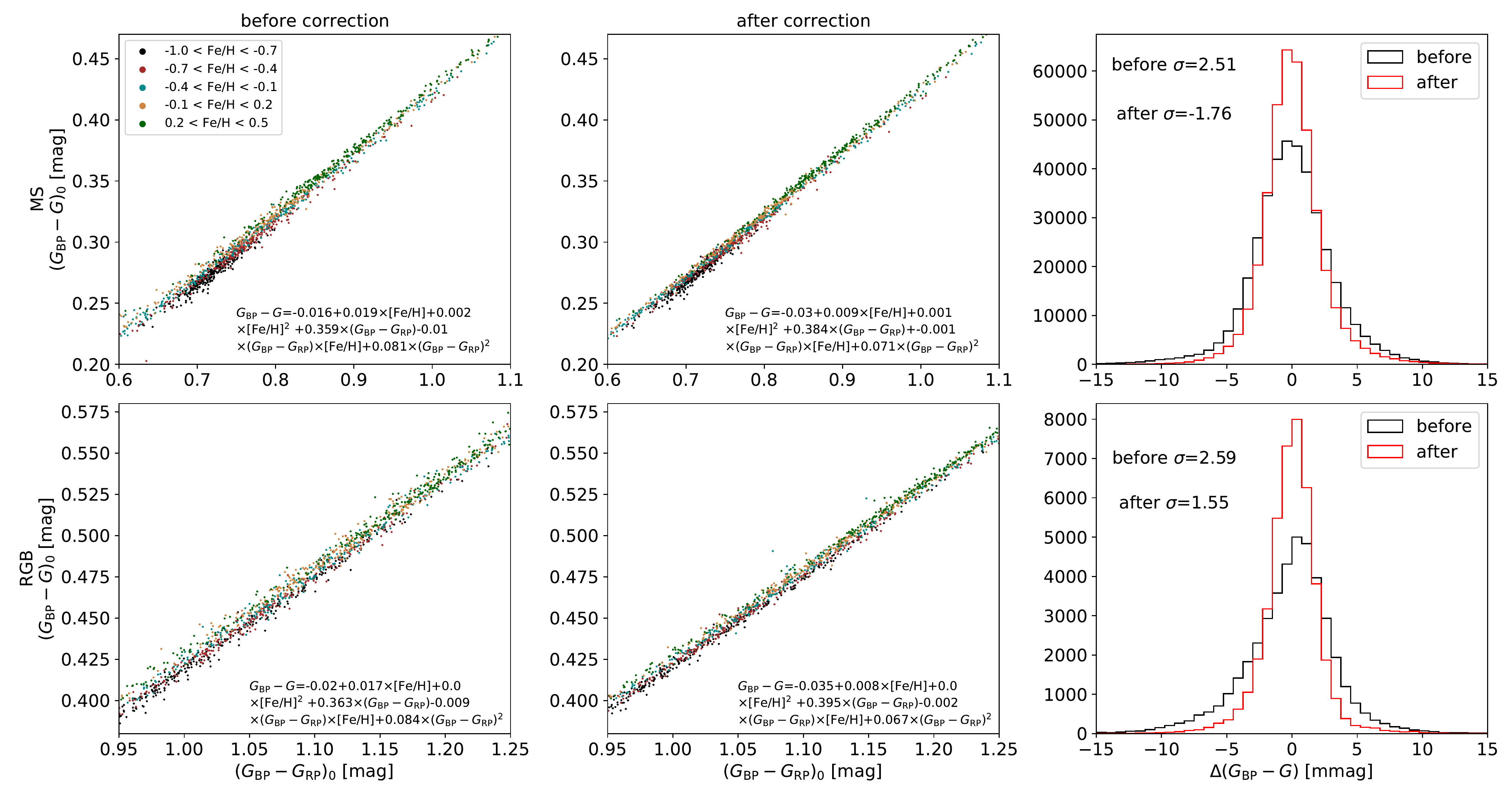}  
    \caption{Left and middle columns compare $G_{\rm BP}-G_{\rm RP}$ vs. $G_{\rm BP}-G$ diagrams before and after corrections, respectively. Right column compares the $G_{\rm BP}-G$ residual distributions before and after corrections. Gaussian fitting is applied, and the fitted $\sigma$ values are texted. \textit{top}: dwarfs; \textit{bottom}: RGB giants.\label{fig:fig9}} 
\end{figure*}

Here we carry out a test of the validity of our calibration curves by comparing the observed and the revised color–color diagrams. The recommended blank solid curves are adopted to correct $G-G_{\rm RP}$ and $G_{\rm BP}-G_{\rm RP}$ for stars of $G$ $>$ 11.5 mag. For stars of $G$ $<$ 11.5 mag, $G_{\rm BP}-G_{\rm RP}$ colors are corrected by colored curves. For stars of $G_{\rm BP}-G_{\rm RP}$ $>$ 1.13 mag, the calibration curve of $G_{\rm BP}-G_{\rm RP}$ = 1.13 mag is used. We correct the whole sample using equation (\ref{eq3}) and compare color-color diagrams before and after correction in the left and middle panels in Figure \ref{fig:fig9}. To exhibit the metallicity-dependent stellar locus clearly, only 300 stars in each metallicity bin are plotted and colored by their metallicities. After correction, the distribution of stars with similar [Fe/H] gets tighter, as expected. Following the work of \citet{yuanloci}, a fourth-order polynomial is adopted to fit the metallicity-dependent stellar loci of the \textit{Gaia} colors. The histograms of the fitting residuals are plotted in the right panels of Figure \ref{fig:fig9}. Both dwarfs and giants show a much smaller $\sigma$ after correction, from 2.51 mmag to 1.76 mmag for dwarfs and from 2.59 mmag to 1.55 mmag for giants. The distribution for giants also becomes more symmetric and Gaussian. 

\begin{figure}[htbp] 
    \centering 
    \includegraphics[width=3.5in]{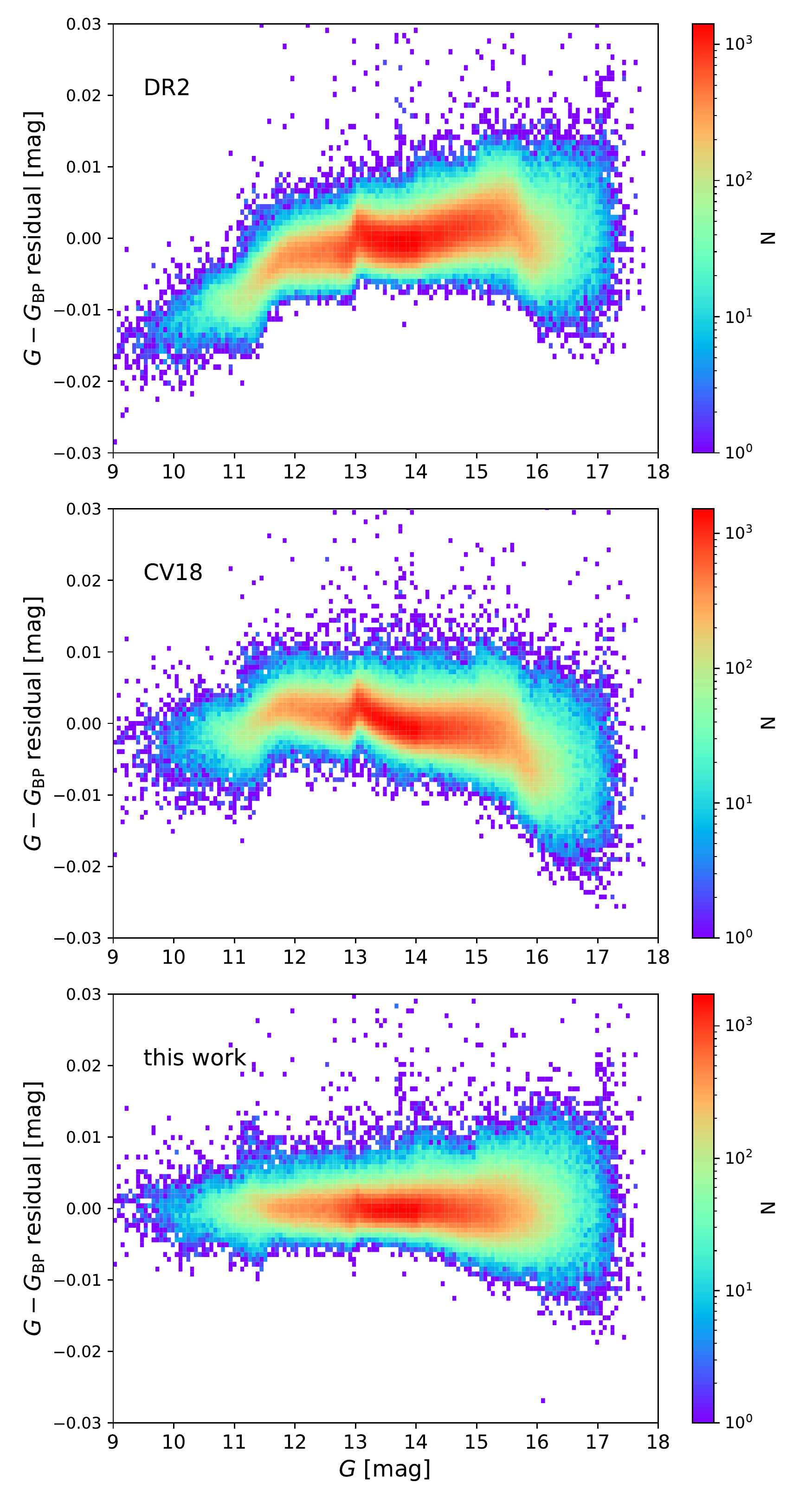}  
    \caption{2D histogram of the $G-G_{\rm BP}$ residual after subtracting the metallicity-dependent color locus. \textit{top}: published DR2 data. \textit{middle}: applying \citetalias{casa2018} corrections. \textit{bottom}: applying corrections from this work. \label{fig14}} 
\end{figure}

Like Figure 31 in \citet{arenou2018}, we also plot a 2D histogram of the $G-G_{\rm BP}$ residuals in Figure \ref{fig14} after subtracting the metallicity-dependent stellar color locus. Subplot labelled DR2 is identical to \citet{arenou2018}, showing a clear trend with G magnitude. After the linear correction in the G magnitude proposed by \citetalias{casa2018}, the overall trend disappears, but still shows small scale features at a few mmag level. Likewise, the liner corrections suggested by \citetalias{wei2018} and \citetalias{maw2018} suffer the same problem. After corrections of this work, the trend is flat everywhere. 

The revised color-color diagram is essential in a number of studies, for instance, detecting peculiar objects or determining reliable metallicities for an enormous and magnitude-limited sample of stars from the \textit{Gaia} photometry (Xu et al., to be submitted). We notice that the $G_{\rm BP}-G$ residual distribution of dwarfs after correction is asymmetric at its wings. There is an excess of stars whose colors are redder compared to those predicted by the metallicity-dependent color locus. This asymmetric features can be well explained by binary stars \citep{yuanbf} and used to estimate binary fractions of a huge number of field stars (Niu et al., to be submitted). 

\section{Summary} \label{sec:sum}

In this work, using about half-million stars with high-quality \textit{Gaia} DR2 photometry and LAMOST DR5 stellar parameters , we apply the SCR method to calibrate the \textit{Gaia} $G-G_{\rm RP}$ and $G_{\rm BP}-G_{\rm RP}$ colors. An unprecedented precision of about 1 mmag is achieved, suggesting the great power of the SCR method in calibrating photometric surveys. 

Magnitude dependent trends are revealed for both the $G-G_{\rm RP}$ and $G_{\rm BP}-G_{\rm RP}$ colors in great details, reflecting changes in instrument configurations. Color dependent trends are found for the $G_{\rm BP}-G_{\rm RP}$ color and for stars brighter than $G$ $\sim$ 11.5 mag. The calibration is up to 20 mmag in general and varies a few mmag/mag. Our results are consistent with previous results but with a much improved precision. Implementations of the color and magnitude corrections are provided. A revised color-color diagram of \textit{Gaia} DR2 is given to demonstrate some potential applications of the corrected mmag photometry. The $G$, $G_{\rm BP}$, and $G_{\rm RP}$ passbands are different in \textit{Gaia} EDR3 in the same way that the $G$ passbands were different between DR1 and DR2. We will extend the work to \textit{Gaia} EDR3 data in a separate paper.

\acknowledgments
We acknowledge the anonymous referee for his/her valuable comments that improve the quality of this paper significantly.
We acknowledge Profs. L. Casagrande and J. Maíz Apellániz for providing data of their work. We greatly thank Huiqin Yang for a careful reading of the manuscript. This work is supported by National Science Foundation of China (NSFC) under grant numbers 11603002, 11988101, and 113300034, National Key Research and Development Program of China (NKRDPC) under grant numbers 2016YFA0400804, 2019YFA0405503, and 2019YFA0405504. 

This work has made use of data products from the Guoshoujing Telescope (the Large Sky Area Multi-Object Fiber Spectroscopic Telescope, LAMOST). LAMOST is a National Major Scientific Project built by the Chinese Academy of Sciences. Funding for the project has been provided by the National Development and Reform Commission. LAMOST is operated and managed by the National Astronomical Observatories, Chinese Academy of Sciences. This work has made use of data from the European Space Agency (ESA) mission
{\it Gaia} (\url{https://www.cosmos.esa.int/gaia}), processed by the {\it Gaia}
Data Processing and Analysis Consortium (DPAC,
\url{https://www.cosmos.esa.int/web/gaia/dpac/consortium}). Funding for the DPAC
has been provided by national institutions, in particular the institutions
participating in the {\it Gaia} Multilateral Agreement.

%





\clearpage

\bibliography{sample63}{}
\bibliographystyle{aasjournal}



\end{document}